\documentclass[seceq]{ptptex}

\usepackage{graphicx}
\usepackage{wrapft}

\title{Spectral Statistics and Luminosity Function of a Hard X-ray
Complete Sample of Brightest AGNs}

\author{K.~\textsc{Shinozaki}$^1,$\footnote{shino@phys.metro-u.ac.jp}
T.~\textsc{Miyaji}$^2,$
Y.~\textsc{Ishisaki}$^1,$
Y.~\textsc{Ueda}$^3,$
Y.~\textsc{Ogasaka}$^4,$
K.~\textsc{Hayashida}$^5,$
H.~\textsc{Awaki}$^6$
}

\inst{
$^1$Experimental Astrophysics Group,
Department of Physics, Tokyo Metropolitan University
Tokyo, Japan\\
$^2$Department of Physics, Carnegie Mellon University\\
$^3$ISAS/JAXA\\
$^4$Department of Astrophysics, Nagoya University\\
$^5$Department of Astrophysics, Osaka University\\
$^6$Department of Physics, Ehime University\\

}

\abst{
Investigating the X-ray luminosity function of AGNs in different redshifts
is very important in understanding growth of super massive black holes in 
centers of galaxies over the cosmological time scale.
As a part of this, we are investigating the statistics of the X-ray spectral 
properties of a complete flux-limited sample of bright AGNs. The purpose
of this investigation is to provide the present-day constraint on the 
evolution of absorbed and unabsorbed AGNs, and to constrain the AGN 
population synthesis model of the total X-ray background.
}

\begin{document}

\maketitle

\begin{figure}[htb]
    \parbox{\halftext}{
      \centerline{\includegraphics[width=6.5cm]
        {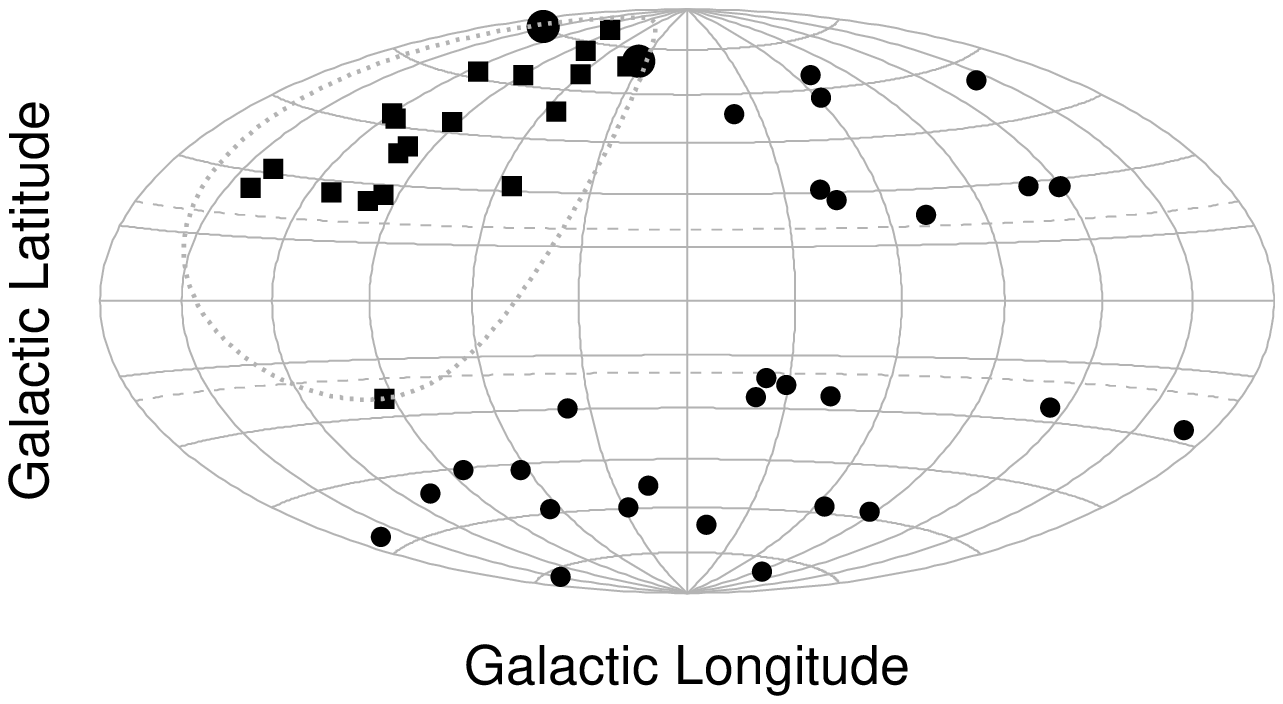}}
\caption{Spatial ditribution of AGNs in our two samples. 
Circles denote the Sample-1 (Piccinotti) and Squares denote
the Sample-2 (Grossan).
Dashed lines correspond to b=$\pm 20^{\circ}$, and Dotted region
represents the selected resion for the sample-2.}
    \label{fig:1}}
    \hfill
    \parbox{\halftext}{
      \centerline{\includegraphics[width=5.0cm]
        {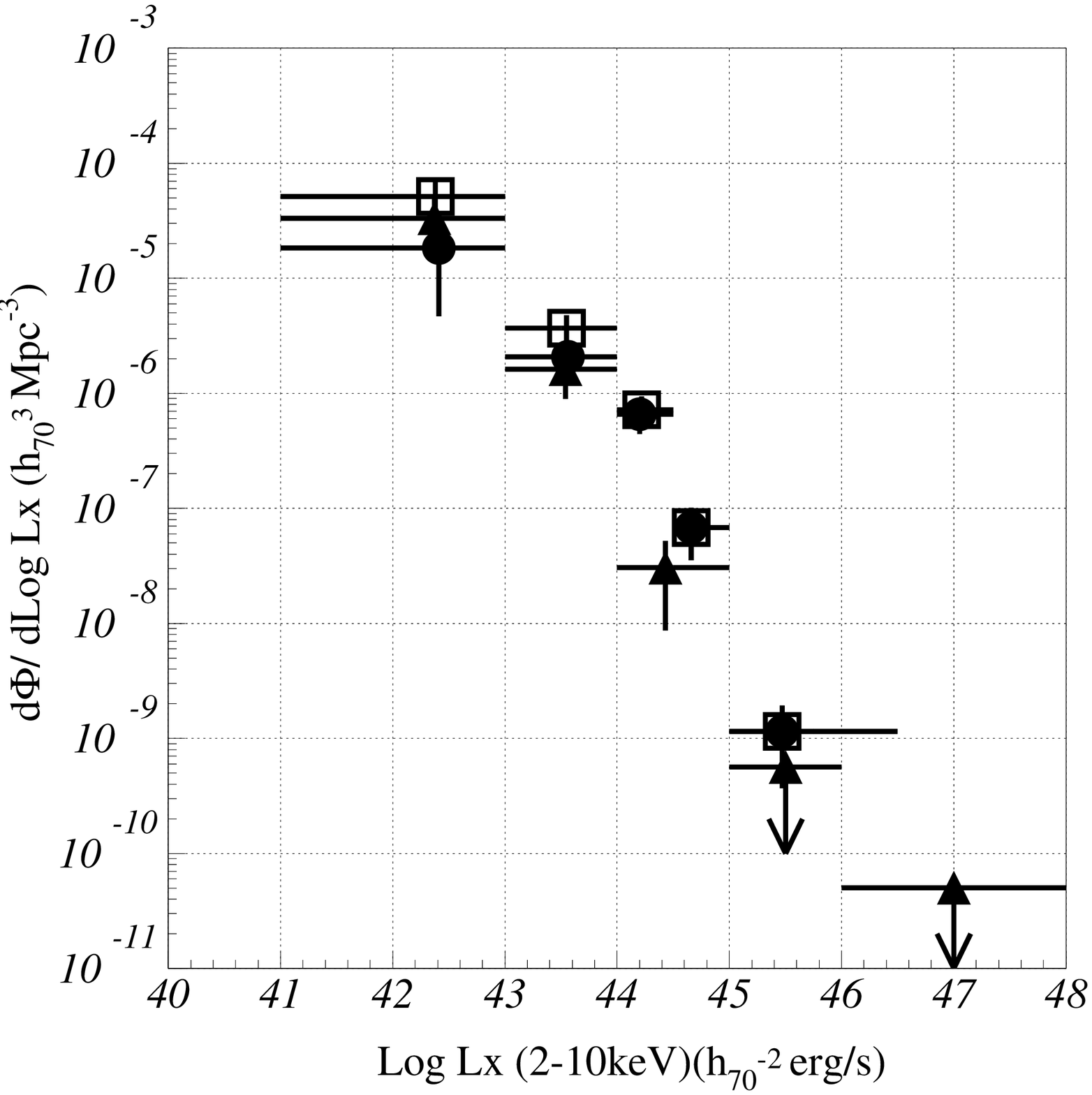}}
      \caption{Hard X-ray luminosity function of all AGNs (Square), X-ray
unabsorbed AGNs (Circle), and absorbed AGNs (Triangle).}
    \label{fig:2}}
\end{figure}

We defined 28 AGNs in Picinotti et al. (1982)\cite{rf:1}, as a Sample-1,
selecting the R15 count rate $>$ 1.25 c s$^{-1}$
$\sim 3 \times 10^{-11}$ erg cm$^{-2}$ s$^{-1}$.
The Galactic latitude $\mid$b$\mid < 20^{\circ}$ 
and the LMC region were excluded.
Two sources which are also in Sample-2 were excluded.
This sample covers 65.5 \% of the sky.
We also defined Sample-2, which contains
20 sources from Grossan (1992)\cite{rf:2}.
This sample was selected from the MC-LASS catalog of HEAO-1 A1/A3
sources and with a flux-limit of  $\sim 1.7 \times 10^{-11}$
erg cm$^{-2}$ s$^{-1}$, inside the 55$^{\circ}$-radius from the NEP.
As a result, we selected 48 sources 
as a complete flux-limited sample with $z <$ 0.4 (Fig.~\ref{fig:1}).
Spectral studies have been made using ASCA (14 sources) or XMM-Newton
(22 sources) observations by ourselves, and other 12 sources were 
derived from literature\cite{rf:3}\tocite{rf:11}.
These two samples are also used to calculate a wide range HXLF in Ueda
et al. (2003)\cite{rf:13}.

\begin{figure}[htb]
    \parbox{\halftext}{
      \centerline{\includegraphics[width=5.0cm]
        {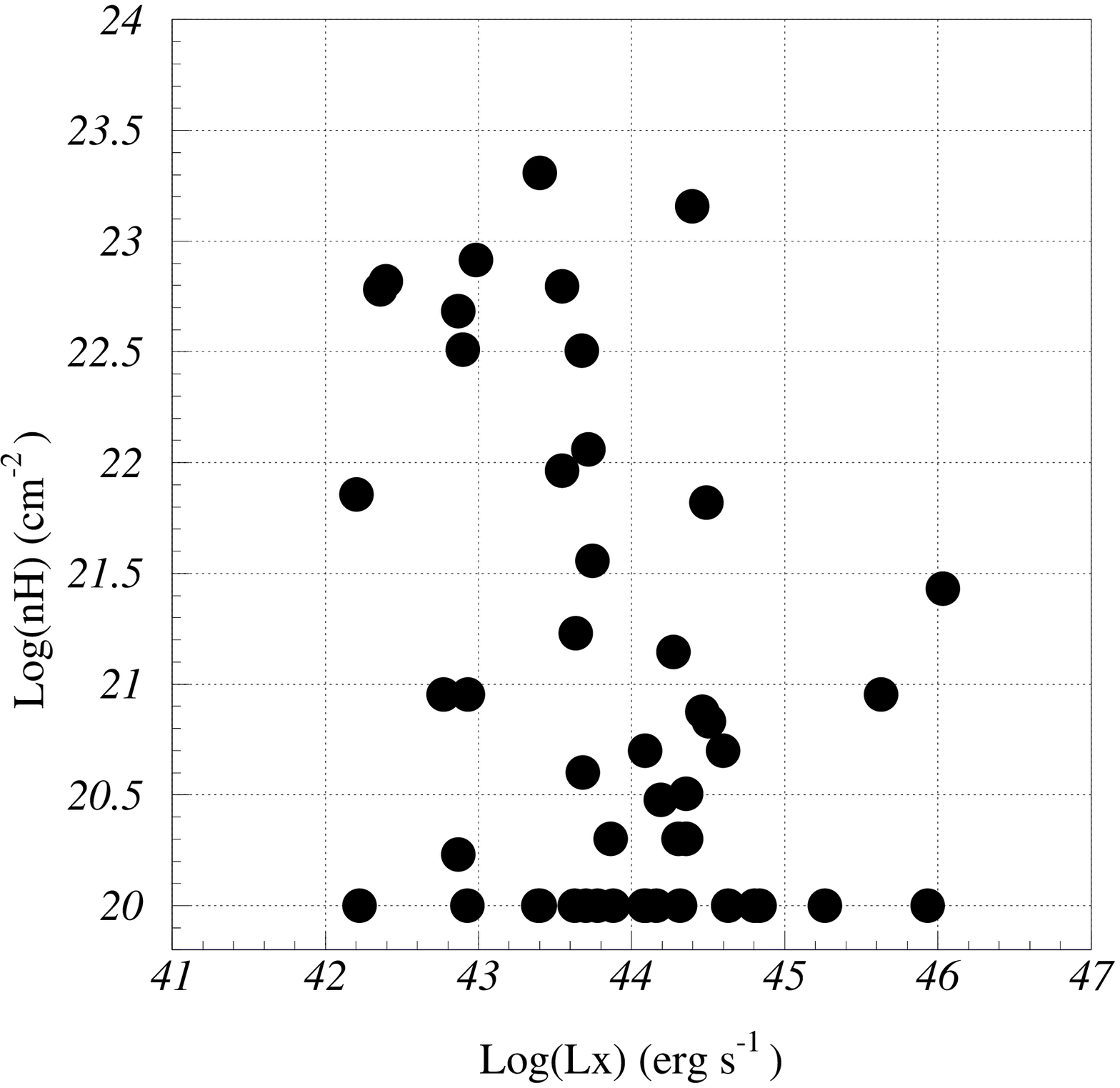}}
\caption{Relation between hard X-ray luminosity and intrinsic $N_{\rm H}$.}
    \label{fig:3}}
    \hfill
    \parbox{\halftext}{
      \centerline{\includegraphics[width=5.0cm]
        {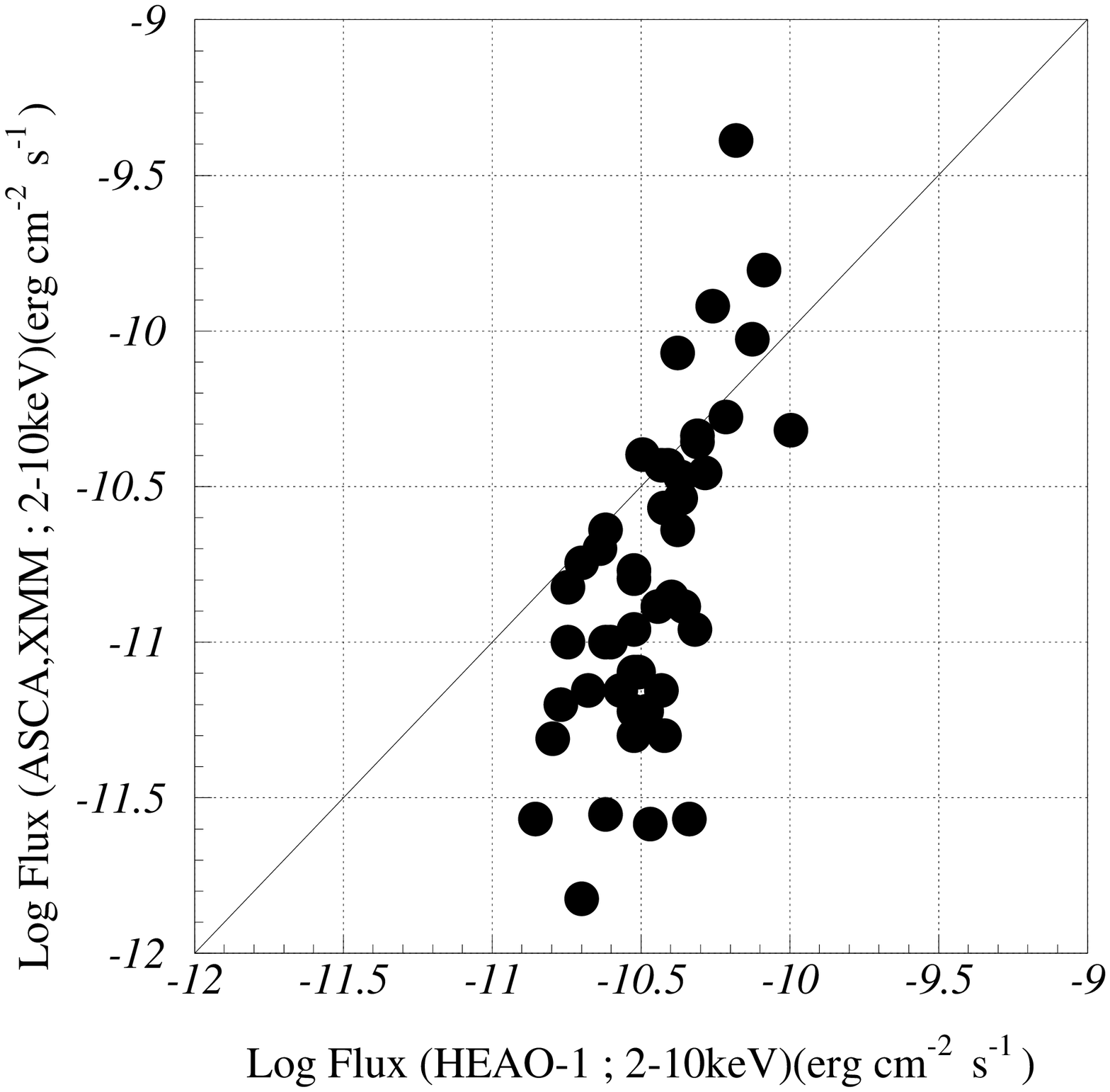}}
      \caption{Comparison of observed flux between ASCA or XMM-Newton, and HEAO-1.}
    \label{fig:4}}
\end{figure}

The spectral measurements enabled us to construct local HXLF for absorbed 
(${\rm Log}\,N_{\rm H}[{\rm cm^{-2}}]> 21.5$) and 
unabsorbed AGNs separately. We calculated each HXLF with the conventional
$\Sigma V_{\rm a}^{-1}$ method\cite{rf:12} (Fig.~\ref{fig:2}).
We found a high luminosity cut-off for the HXLF of the absorbed AGNs
compared with those of unabsorbed AGNs and all AGNs.
In the $L_{\rm x}-N_{\rm H}$ plot (Fig.~\ref{fig:3}), there are no AGNs
in the region where ${\rm Log}\,L_{\rm X} [\rm erg/s]> 44.5$ and 
${\rm Log}\,N_{\rm H}[\rm cm^{-2}]> 21.5$. We would have 9-10 AGNs
in this region if the shapes of the absorbed and unabsorbed AGN HXLF
were identical.
We found that the observed flux with ASCA or XMM-Newton was smaller
than that with HEAO-1 by a factor of 0.42 on average, which was expected
for re-observation of sources with a variability-amplitude of a factor 
$\sim$2.6 (Fig.~\ref{fig:4}).

%

\end{document}